\documentclass[fleqn,usenatbib]{mnras}

\usepackage{newtxtext,newtxmath}
\usepackage{relsize, textcmds}

\usepackage[T1]{fontenc}

\DeclareRobustCommand{\VAN}[3]{#2}
\let\VANthebibliography\thebibliography
\def\thebibliography{\DeclareRobustCommand{\VAN}[3]{##3}\VANthebibliography}

\usepackage{graphicx, caption, subcaption}	
\usepackage{amsmath}	
\usepackage{lineno}

\title[BALs in Early DESI Data]{Broad Absorption Line Quasars in the Dark Energy Spectroscopic Instrument Early Data Release}

\author[Filbert et al.]{
\parbox{\textwidth}{
\Large
S.~Filbert,$^{1,2,3,4}$
P.~Martini,$^{1,2,3}$
K.~Seebaluck,$^{1,2,3}$
L.~Ennesser,$^{2,3}$
D.~M.~Alexander,$^{5,6}$
A.~Bault,$^{7}$
A.~Brodzeller,$^{4}$
H.~K.~Herrera-Alcantar,$^{8}$
P.~Montero-Camacho,$^{9}$
I.~P\'erez-R\`afols,$^{10}$
C.~Ram\'irez-P\'erez,$^{11}$
C.~Ravoux,$^{12,13}$
T.~Tan,$^{14}$
J.~Aguilar,$^{15}$
S.~Ahlen,$^{16}$
S.~Bailey,$^{15}$
D.~Brooks,$^{17}$
T.~Claybaugh,$^{15}$
K.~Dawson,$^{4}$
A.~de la Macorra,$^{18}$
P.~Doel,$^{17}$
K.~Fanning,$^{2,3}$
A.~Font-Ribera,$^{11}$
J.~E.~Forero-Romero,$^{19,20}$
S.~Gontcho A Gontcho,$^{15}$
J.~Guy,$^{15}$
D.~Kirkby,$^{7}$
A.~Kremin,$^{15}$
C.~Magneville,$^{13}$
M.~Manera,$^{21,11}$
A.~Meisner,$^{22}$
R.~Miquel,$^{23,11}$
J.~Moustakas,$^{24}$
J.~Nie,$^{25}$
W.~J.~Percival,$^{26,27,28}$
F.~Prada,$^{29}$
M.~Rezaie,$^{30}$
G.~Rossi,$^{31}$
E.~Sanchez,$^{32}$
M.~Schubnell,$^{33,34}$
H.~Seo,$^{35}$
G.~Tarl\'{e},$^{34}$
B.~A.~Weaver,$^{22}$
and Z.~Zhou$^{25}$
}
\vspace{0.4cm}
\\
\parbox{\textwidth}{
\scriptsize
$^{1}$ Department of Astronomy, The Ohio State University, 140 W 18th Avenue, Columbus, OH 43210, USA\\
$^{2}$ Center for Cosmology and AstroParticle Physics, The Ohio State University, 191 West Woodruff Avenue, Columbus, OH 43210, USA\\
$^{3}$ Department of Physics, The Ohio State University, 191 West Woodruff Avenue, Columbus, OH 43210, USA\\
$^{4}$ Department of Physics and Astronomy, The University of Utah, 115 South 1400 East, Salt Lake City, UT 84112, USA\\
$^{5}$ Centre for Extragalactic Astronomy, Department of Physics, Durham University, South Road, Durham, DH1 3LE, UK\\
$^{6}$ Institute for Computational Cosmology, Department of Physics, Durham University, South Road, Durham DH1 3LE, UK\\
$^{7}$ Department of Physics and Astronomy, University of California, Irvine, 92697, USA\\
$^{8}$ Departamento de F\'{i}sica, Universidad de Guanajuato - DCI, C.P. 37150, Leon, Guanajuato, M\'{e}xico\\
$^{9}$ Department of Astronomy, Tsinghua University, 30 Shuangqing Road, Haidian District, Beijing, China, 100190\\
$^{10}$ Departament de F\'{\i}sica Qu\`{a}ntica i Astrof\'{\i}sica, Universitat de Barcelona, Mart\'{\i} i Franqu\`{e}s 1, E08028 Barcelona, Spain\\
$^{11}$ Institut de F\'{i}sica d’Altes Energies (IFAE), The Barcelona Institute of Science and Technology, Campus UAB, 08193 Bellaterra Barcelona, Spain\\
$^{12}$ Aix Marseille Univ, CNRS/IN2P3, CPPM, Marseille, France\\
$^{13}$ IRFU, CEA, Universit\'{e} Paris-Saclay, F-91191 Gif-sur-Yvette, France\\
$^{14}$ Sorbonne Universit\'{e}, CNRS/IN2P3, Laboratoire de Physique Nucl\'{e}aire et de Hautes Energies (LPNHE), FR-75005 Paris, France\\
$^{15}$ Lawrence Berkeley National Laboratory, 1 Cyclotron Road, Berkeley, CA 94720, USA\\
$^{16}$ Physics Dept., Boston University, 590 Commonwealth Avenue, Boston, MA 02215, USA\\
$^{17}$ Department of Physics \& Astronomy, University College London, Gower Street, London, WC1E 6BT, UK\\
$^{18}$ Instituto de F\'{\i}sica, Universidad Nacional Aut\'{o}noma de M\'{e}xico,  Cd. de M\'{e}xico  C.P. 04510,  M\'{e}xico\\
$^{19}$ Departamento de F\'isica, Universidad de los Andes, Cra. 1 No. 18A-10, Edificio Ip, CP 111711, Bogot\'a, Colombia\\
$^{20}$ Observatorio Astron\'omico, Universidad de los Andes, Cra. 1 No. 18A-10, Edificio H, CP 111711 Bogot\'a, Colombia\\
$^{21}$ Departament de F\'{i}sica, Serra H\'{u}nter, Universitat Aut\`{o}noma de Barcelona, 08193 Bellaterra (Barcelona), Spain\\
$^{22}$ NSF's NOIRLab, 950 N. Cherry Ave., Tucson, AZ 85719, USA\\
$^{23}$ Instituci\'{o} Catalana de Recerca i Estudis Avan\c{c}ats, Passeig de Llu\'{\i}s Companys, 23, 08010 Barcelona, Spain\\
$^{24}$ Department of Physics and Astronomy, Siena College, 515 Loudon Road, Loudonville, NY 12211, USA\\
$^{25}$ National Astronomical Observatories, Chinese Academy of Sciences, A20 Datun Rd., Chaoyang District, Beijing, 100012, P.R. China\\
$^{26}$ Department of Physics and Astronomy, University of Waterloo, 200 University Ave W, Waterloo, ON N2L 3G1, Canada\\
$^{27}$ Perimeter Institute for Theoretical Physics, 31 Caroline St. North, Waterloo, ON N2L 2Y5, Canada\\
$^{28}$ Waterloo Centre for Astrophysics, University of Waterloo, 200 University Ave W, Waterloo, ON N2L 3G1, Canada\\
$^{29}$ Instituto de Astrof\'{i}sica de Andaluc\'{i}a (CSIC), Glorieta de la Astronom\'{i}a, s/n, E-18008 Granada, Spain\\
$^{30}$ Department of Physics, Kansas State University, 116 Cardwell Hall, Manhattan, KS 66506, USA\\
$^{31}$ Department of Physics and Astronomy, Sejong University, Seoul, 143-747, Korea\\
$^{32}$ CIEMAT, Avenida Complutense 40, E-28040 Madrid, Spain\\
$^{33}$ Department of Physics, University of Michigan, Ann Arbor, MI 48109, USA\\
$^{34}$ University of Michigan, Ann Arbor, MI 48109, USA\\
$^{35}$ Department of Physics \& Astronomy, Ohio University, Athens, OH 45701, USA\\
}
}

\pubyear{2023}

\begin{document}
\label{firstpage}
\pagerange{\pageref{firstpage}--\pageref{lastpage}}
\maketitle

\begin{abstract}
Broad absorption line (BAL) quasars are characterized by gas clouds that absorb flux at the wavelength of common quasar spectral features, although blueshifted by velocities that can exceed $0.1c$. BAL features are interesting as signatures of significant feedback, yet they can also compromise cosmological studies with quasars by distorting the shape of the most prominent quasar emission lines, impacting redshift accuracy  and measurements of the matter density distribution traced by the Lyman-$\alpha$ forest. We present a catalog of BAL quasars discovered in the Dark Energy Spectroscopic Instrument (DESI) survey Early Data Release, which were observed as part of DESI Survey Validation, as well as the first two months of the main survey. We describe our method to automatically identify BAL quasars in DESI data, the quantities we measure for each BAL, and investigate the completeness and purity of this method with mock DESI observations. We mask the wavelengths of the BAL features and re-evaluate each BAL quasar redshift, finding new redshifts which are $243\,{\rm km}\,{\rm s}^{-1}$ smaller on average for the BAL quasar sample. These new, more accurate redshifts are important to obtain the best measurements of quasar clustering, especially at small scales. Finally, we present some spectra of rarer classes of BALs that illustrate the potential of DESI data to identify such populations for further study.

\end{abstract}




\section{Introduction}

Quasars have been a valuable cosmological tool since the realization by \citet{schmidt63} that they were outside the local universe, and therefore had the potential to probe early into cosmic history. In the intervening decades, quasars have become a standard tool for cosmology at redshifts above $z \sim 1.5$. At these redshifts, the space density of luminous quasars is much higher than in the local universe \citep[e.g.][]{hopkins07}. Their higher space density combined with their high luminosity make quasars a valuable probe of large-scale structure.

The Sloan Digital Sky Survey \citep[SDSS,][]{york00} and especially its extensions to SDSS-III \citep{eisenstein11} and SDSS-IV \citep{blanton17} have targeted large numbers of quasars to study their physical properties, their evolution, and use them as probes of large-scale structure. The SDSS-III Baryon Oscillation Spectroscopic Survey \citep[BOSS,][]{dawson13} and its SDSS-IV successor eBOSS \citep{dawson16} measured the baryon acoustic oscillation feature with quasars as point tracers of the matter-density field \citep{neveux20,hou21}, through studies of the matter distribution in the Lyman-$\alpha$ (Ly$\alpha$) forest \citep{busca13,bautista17,dumasdesbourboux20, desainteagathe19}, and by cross-correlating quasars with Ly$\alpha$ forest absorption \citep{fontribera19, blomqvist19}. 

A complication to these studies is that 10--30\% of quasars exhibit broad absorption line (BAL) troughs \citep{foltz90,trump06}. The BAL class of quasars was first defined by \citet{weymann91} as quasars with troughs blueshifted by at least $3000\,{\rm km}\,{\rm s}^{-1}$ and absorption velocity widths of at least $2000\,{\rm km}\,{\rm s}^{-1}$. BAL troughs are most commonly observed associated with the prominent \ion{C}{IV} emission line, although are also observed associated with many other high-ionization features, such as \ion{Si}{IV} and \ion{N}{V}. Quasars with absorption associated with high-ionization features are dubbed HiBALs. Quasars more rarely exhibit absorption associated with lower-ionization features such as \ion{Mg}{II} and \ion{Fe}{II}, in which case they are called LoBALs, and FeLoBALs when absorption from excited states of \ion{Fe}{II} and \ion{Fe}{III} are present \citep{hall02, trump06}. 

BAL features are a complication for cosmological studies for two reasons. The first is that BAL features can significantly distort the shape of the main emission features that are used to determine the redshift of the quasar, such as the \ion{C}{IV} emission line. This can confuse and bias algorithms that automatically determine quasar redshifts for large surveys. Specifically, BAL troughs can to lead to an overestimate of the quasar redshift because the absorption can depress the blue wing of one or more spectral features. Secondly, BAL features may also be associated with many of the emission features in the Ly$\alpha$ forest region used for cosmological studies. Highly blueshifted BAL features associated with prominent lines like Ly$\alpha$ and \ion{N}{V}, as well as weaker lines such as \ion{P}{V} and \ion{S}{IV} \citep{masribas19}, can produce absorption that is often indistinguishable from the neutral hydrogen absorption that comprises the Ly$\alpha$ forest. 

These complications have made it critically important to identify BALs for large-scale structure studies using quasars. With large surveys such as SDSS and now DESI, the use of automated methods to identify and classify BAL quasars is more important than ever. \cite{reichard03} were the first to use an automated method to identify BAL quasars in the SDSS early data, presenting 224 visually confirmed BAL quasars. The SDSS DR12 quasar catalog contains 297,301 quasars, 29,580 of which exhibited BAL features \citep{paris17}. This is 13\% of all quasars above $z = 1.57$, as it is only above this redshift that BAL features could be unambiguously identified on the blue side of the \ion{C}{IV} emission line. Given the size of these quasar samples, subsequent quasar samples from SDSS DR14 \citep{paris18} and DR16 \citep{lyke20} used a variety of automated algorithms, including machine-learning methods \citep{guo19}. 

The Dark Energy Spectroscopic Instrument (DESI) survey plans to measure about 2.8 million quasars, including over 800,000 at $z>2.1$ that will probe the Ly$\alpha$ forest \citep{chaussidon23}. This further increases the importance of developing automated methods to identify BALs, quantify their impact on quasar redshift errors, and further study how to optimally include BAL quasars in cosmological analyses. This is important because BAL quasars had historically been excluded from Ly$\alpha$ studies \citep[e.g.][]{bautista17,dumasdesbourboux20}, which removes $10-15$\% of the sample. \citet{ennesser22} recently studied the impact of BALs on Ly$\alpha$ forest studies in detail and developed criteria to maintain the vast majority of BAL quasars in Ly$\alpha$ cosmology measurements. They showed that after masking the expected locations of BAL features in the Ly$\alpha$ forest continuum region, they could obtain cosmological results that were consistent with non-BAL quasars. Including BALs after masking absorption features produced a pronounced improvement in the fractional errors in the correlation function both by increasing the total number of sightlines with Ly$\alpha$ measurements and because the BAL quasar spectra had on average higher signal-to-noise ratio (SNR). 

In this paper we describe the algorithm that we plan to use to identify BAL quasars in the DESI survey, as well as present a catalog of the BAL quasars from Early Data Release (EDR) observations and the first two months of main survey operations (M2). Section~\ref{sec:data} has a description of the DESI observations, how we identify BALs, and how we add BALs to mock DESI datasets. In Section~\ref{sec:cp} we investigate the completeness and purity of the BAL identification algorithm with DESI mock data. We then present the DESI EDR+M2 BAL catalog in Section~\ref{sec:balcat}. This section includes a description of the catalog data model as well as statistical properties of the BAL quasars. The catalog contains information about individual BALs, including the locations of all of the absorption troughs associated with \ion{C}{IV}, as the velocity offsets of these troughs are critical to masking the features that impact both redshifts and the forest. We then study the impact of BAL features on quasar redshift measurements in Section~\ref{sec:balz}. We quantify the difference in redshifts with and without masking the BAL features, and present new redshifts for the BAL quasars based on our masking approach. In Section~\ref{sec:unusualbals} we mention some unusual BALs and we conclude in Section~\ref{sec:summary} with a summary of our main results. 

\section{Data} \label{sec:data}

The BALs in this paper were discovered in observations during DESI Survey Validation (SV), which are described in \citet{svpaper}, and the first two months of the main survey (M2). SV extended from December 2020 through mid-May 2021 and M2 from mid-May to mid-July 2021. The purposes of Survey Validation were to test the data quality, optimize target selection, and exercise the operational and analysis programs. The two specific periods of SV included here are Target Selection Validation (also referred to as SV1) and the One-Percent Survey (also referred to as SV3). The SV observations are publicly available as part of the DESI Early Data Release (EDR), which is described in \citet{edrpaper}. In this section we briefly present some background information about DESI observations, describe our procedure to identify BALs in these data, and finally describe the mock datasets we use to verify our approach. We constructed separate BAL catalogs for SV1, SV3, and M2. Unless otherwise specified, we perform our analysis with the combination of all three of these catalogs. 

\subsection{DESI Observations} \label{sec:obs}

\begin{figure*}
    \includegraphics[width=7in]{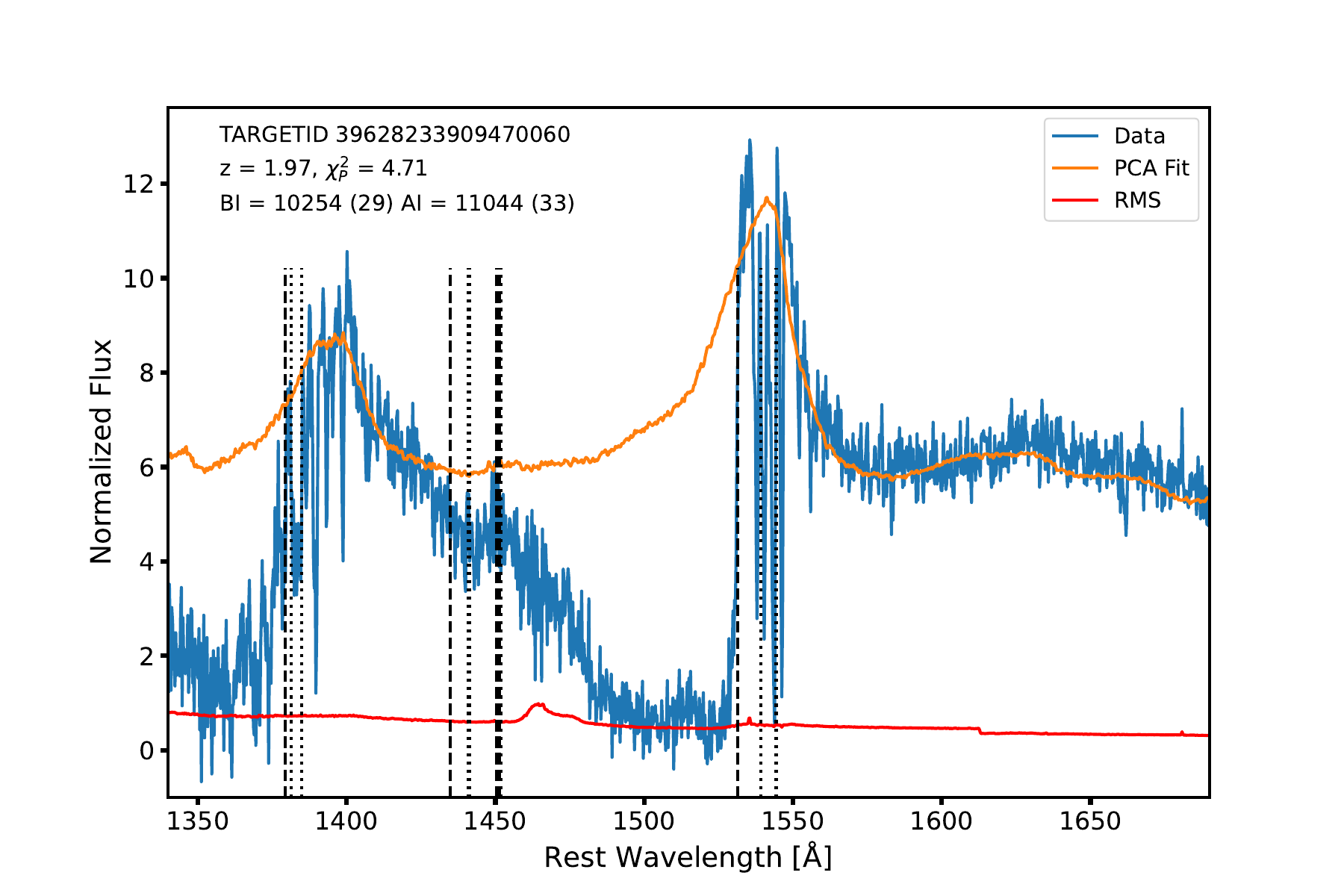}
    \caption{Spectrum of a BAL quasar from early DESI data. The $\chi^2_P$ represents the goodness of fit of the PCA fit ({\it orange lines}) to the data ({\it blue line}) with RMS flux uncertainty({\it red line}) for regions of the spectra where absorption is not present. The dotted lines delimit multiple absorption troughs meeting the AI criterion, whereas the thicker dashed black lines correspond to BI absorption. The values of AI and BI absorption are shown in the legend, along with their corresponding errors. These absorption definitions and their errors are in Equations \ref{eq:BI}, \ref{eq:BI_err}, \ref{eq:AI}, and \ref{eq:AI_err}.}
    \label{fig:bal_spectra}
\end{figure*}

The goal of DESI is to study cosmic acceleration with a spectroscopic survey of 40 million galaxies and quasars in just five years \citep{desi16a}. The DESI collaboration plans to use these data to measure distances from the nearby universe to beyond $z > 3.5$ with the baryon acoustic oscillation method, as well as employ redshift-space distortions to measure the growth of cosmic structures and test potential modifications to general relativity. The DESI collaboration is conducting this survey with a highly multiplexed fiber-fed spectrograph at the Mayall 4\,m telescope of the Kitt Peak National Observatory \citep{desi22}. This instrument has a $3\deg$ diameter field of view that was designed to observe 5000 targets in a single observation. The light from each target is fed into one of ten bench-mounted spectrographs that are located in a climate-controlled room. Each of these spectrographs records light from $360 - 980$\,nm split among three distinct wavelength channels. The blue channel is especially important for the quasars at $z > 2.1$ that are used to study the Ly$\alpha$ forest. This channel extends from $360 - 593$\,nm and ranges in resolution from about $2000-3500$. The neighboring red channel covers $560-772$\,nm with a resolution of about $3500-4500$ and the near-infrared channel covers $747-980$\,nm with a resolution of about $4000-5000$. More details of the instrumentation are described in \citet{desi16b} for the technical design, \citet{silber23} for the focal plane system, and \citet{miller23} for the optical corrector and support system. 

DESI target selection \citep{myers22} includes both quasars at $z > 0.9$ that are used as direct tracers of the dark matter distribution and $z > 2.1$ quasars that are used to trace the matter distribution in the Ly$\alpha$ forest \citep{chaussidon23}. This selection is based on significant imaging from the DESI Legacy Imaging Surveys \citep{zou17,dey19}. DESI observes 310 quasar targets per deg$^{2}$ and successfully identifies more than 200 quasars per deg$^{2}$, including over 60 per deg$^{2}$ at $z>2.1$ \citep{chaussidon23}. As they are lower surface density than other targets, the quasars are assigned higher priority for fiber assignment during dark time than the other dark time targets (Luminous Red Galaxies and Emission Line Galaxies). Some fibers are devoted to standard stars for flux calibration and to empty regions to measure and subtract the spectrum of the night sky. DESI is able to achieve superb sky subtraction and flux calibration due to a combination of excellent stability and sophisticated data processing algorithms. One-dimensional spectral extractions, redshifts, and spectral classifications are typically available the morning after the observations. \citet{guy23} describe the DESI data processing pipeline in detail and \citet{schlafly23} describe survey operations. The software for target selection, the spectroscopic pipeline, and survey operations is publicly available.\footnote{https://github.com/desihub}

The DESI collaboration constructs quasar catalogs from a combination of three classification routines: Redrock\footnote{https://github.com/desihub/redrock} (S. Bailey et al. 2023, in preparation), QuasarNET \citep{busca18, farr20b}, and the \ion{Mg}{II} afterburner. Redrock performs a $\chi^2$ analysis as a function of redshift for a range of spectral templates developed by \citet{brodzeller23} to identify the best redshift and spectral classification and correctly classifies the vast majority of quasars \citep{alexander23}. Both QuasarNET and the \ion{Mg}{II} afterburner are used to check quasar targets that were not classified as quasars by Redrock. QuasarNET is a machine learning algorithm that employs convolutional neural networks for both classification and redshift estimation. The \ion{Mg}{II} afterburner searches for broad \ion{Mg}{II} emission at the best-fitting Redrock redshift and reclassifies the object as a quasar if \ion{Mg}{II} emission is significant, regardless of the spectral type initially provided by Redrock \citep{chaussidon22}. \citet{chaussidon23} further describe the quasar target selection procedure and \citet{alexander23} describe the visual inspection process that we employed to quantify the performance of the redshift and classification algorithms. The quasar catalogs for SV1, SV3, and M2 include all objects classified as quasars by these algorithms.

\subsection{BAL Identification} \label{sec:identify}

There are a several methods used to quantify the strength of absorption troughs in the spectra of BAL quasars. The first, and in some sense the strictest definition, is the Balnicity Index (BI) proposed by \citet{weymann91}. This definition is especially useful for characterizing quasars that have strong, broad absorption from one or more continuous high column density clouds with a large velocity ($>3000\,\mathrm{km\,s}^{-1}$) relative to the quasar. The equation for BI is:

\begin{equation}\label{eq:BI}
    BI = - \int_{25000}^{3000} \left[ 1- \frac{f(v)}{0.9} \right] C(v) dv ~\rm{.}
\end{equation}

The term $f(v)$ is the normalized flux density of the quasar, which is the observed quasar spectral energy distribution divided by a model fit to the quasar that does not have BAL features. The integration variable $v$ is the velocity displacement blueward of the line centroid, which is usually \ion{C}{IV}. In some cases, BI is also calculated relative to \ion{Si}{IV}, \ion{Mg}{II}, and other lines. $C$ is a constant that is zero unless the term $[1 - \frac{f(v)}{0.9}]$ is greater than zero for more than $2000\,\mathrm{km\,s}^{-1}$, in which case it is set to one. The result is that a trough will have a BI value greater than zero only after the quasar flux is more than 10\% below the estimated continuum flux for a contiguous span of at least $2000\,\mathrm{km\,s}^{-1}$. 

We estimate the error in BI for each BAL quasar classified using the method introduced by \cite{trump06}, although modified by \citet{guo19} to include the uncertainty in the continuum fitting from principal component analysis (PCA). The equation used for the error in BI is:

\begin{equation}\label{eq:BI_err}
    \sigma_{BI}^2 = - \int_{25000}^{3000} \left(
    \frac{\sigma_{f(v)}^{2}  + \ \sigma_{PCA}^{2}}{(0.9)^{2}} \right) C(v) dv ~\rm{.}
\end{equation}

\noindent
where $\sigma_{f(v)}^{2}$ is the variance in the normalized flux density and $\sigma_{PCA}^{2}$ is the variance of the PCA fit. 

Although BI is a robust estimator of absorption, it misses many BAL features because it starts $3000 \,\mathrm{km\,s}^{-1}$ blueward of the line center and is insensitive to features that are narrower than $2000\,\mathrm{km\,s}^{-1}$. The absorption index (AI), introduced by \citet{hall02}, is an alternate estimator of the absorption strength. Unlike BI, the presence of absorption is evaluated up to the line centroid. Additionally, $C(v)$ is set to one after the trough extends more than $450\,\mathrm{km\,s}^{-1}$, rather than $2000\,\mathrm{km\,s}^{-1}$ for BI. The equations for AI and the corresponding uncertainty are: 

\begin{equation}\label{eq:AI}
    AI = - \int_{25000}^{0} \left[ 1-\frac{f(v)}{0.9} \right] C(v) dv
\end{equation}
and
\begin{equation}\label{eq:AI_err}
    \sigma_{AI}^2 = - \int_{25000}^{0} \left(
    \frac{\sigma_{f(v)}^{2}  + \ \sigma_{PCA}^{2}}{(0.9)^{2}} \right) C(v) dv ~\rm{.}
\end{equation}

The AI criterion captures about $4-5$ times more BALs than the BI criterion, and most importantly it captures BAL features that could compromise precise cosmological measurements. We therefore consider any object with $AI > 0$ to be a BAL quasar, although we also calculate BI for comparison with previous work. Furthermore, we only consider absorption from \ion{C}{IV} to classify a quasar as a BAL quasar. We consequently only search for BALs in quasars where the observed spectra extend to $25000\,\mathrm{km\,s}^{-1}$ blueward of the \ion{C}{IV} line. We also require that the spectra extend redward of \ion{C}{IV} to at least 1633\,\AA\ in order to accurately fit the quasar spectra with PCA components, which effectively sets an upper limit in redshift. Because of these two constraints, we only look for BALs in quasar spectra between $1.57 < z < 5.0$. This differs slightly from the implementation used by \citet{guo19} which searched for absorption in the redshift range between 1.57 and 5.56. As we only conduct our search around the high ionization \ion{C}{IV} line, all of the BALs appear to be high-ionization BALs or HiBALs; however, as we only search in this wavelength range we do not classify the BALs into categories such as HiBALs, LoBALs (low ionization BALs), and FeLoBALs \citep[LoBALs with Fe emission;][]{trump06,hall02,gibson09}. We briefly discuss the prospects for identifying other types of BALs in Section~\ref{sec:unusualbals}.

We calculate AI and BI, and other BAL-specific quantities (as in Table~\ref{tab:bal_cols}) with software in the \texttt{baltools}\footnote{https://github.com/paulmartini/baltools} repository. We refer to the code that identifies BALs and measures their properties as the \texttt{balfinder}. The \texttt{balfinder} code was originally developed for SDSS and used a convolution neural network (CNN), as described in \citet{guo19}. The implementation of the \texttt{balfinder} for DESI does not use a CNN, and instead relies on PCA to fit spectra for a given quasar from the five eigenspectra in \citet{guo19}. This PCA fit to each spectrum is iterative, in that any regions of the spectrum with BAL features according to Equation~\ref{eq:AI} are masked before the next iteration. This continues for ten iterations of fitting and masking. Figure~\ref{fig:bal_spectra} shows an example PCA fit for a DESI quasar.  

The best-fit PCA representation is used to estimate the quasar continuum if no absorption were present, and is used to compute the normalized flux density $f(v)$ in the equations for AI and BI. Each quasar may have multiple troughs that satisfy the AI and/or BI criteria, and we refer to the number of such troughs as $N_{450}$ and $N_{2000}$, respectively. The code also records the minimum and maximum velocity for each trough. The sum of the absorption troughs that meet the AI and BI criteria are the AI and BI values for the quasar. All of these data are stored for each BAL quasar, as well as other properties such as the coefficients for each of the eigenspectra used in the PCA fits, the errors $\sigma_{AI}$ and $\sigma_{BI}$, and the $\chi^2$ of the PCA fit. The BAL properties included in the catalog are summarized in Table~\ref{tab:bal_cols}.

In some rare cases the PCA fitting leads to mis-identifications. Based on our visual inspection of hundreds of BAL quasars, this occurs because the PCA components are unable to fit the full range of quasar diversity. Approximately $15\%$ of objects identified as BALs by the \texttt{balfinder} which have bad $\chi^2_{PCA}$ fits were not initially classified as quasars by Redrock, suggesting that these objects already have strange spectra, and were re-classified as quasars by the afterburners discussed at the end of Section \ref{sec:obs}. In any case, should the initial PCA fit be relatively poor, the \texttt{balfinder} may incorrectly identify and mask non-BAL regions, and this masking may compound the misclassification. Based on visual inspection, fits with $\chi^2_{PCA} > 10$ are suspect. These high $\chi^2_{PCA}$ values are uncommon, comprising only about 2\% of the entire BAL catalog. The $\chi^2_{PCA}$ quantity is in the BAL catalog as \q{PCA\_CHI2}. 

The presence of high $\chi^2_{PCA}$ fits hint at a failure of the templates discussed in Section \ref{sec:mocks} to reproduce quasar spectral diversity. This is one limitation of using mock spectra created using these PCA templates to estimate completeness and purity of the BAL quasar sample as it is unclear whether these templates span the diversity of BAL quasars observed by DESI. Section \ref{sec:mocks} further details the creation of the mock spectra used to estimate the purity and completeness of the BAL quasar sample.  



\begin{table*}
    \centering
    \begin{tabular}{|p{1cm}||p{3cm}|p{1.5cm}|p{10.5cm}|}
         \hline
         \multicolumn{4}{|c|}{DESI BAL Catalog Columns} \\
         \hline
         Column & Name & Data Type & Description\\
         \hline
         
65  & PCA\_COEFFS          & FLOAT[5]  & Coefficients on PCA       eigenbasis for spectrum fit.\\ 

66  & PCA\_CHI2            & FLOAT     & Goodness of fit from      PCA components. \\ 

67  & BAL\_PROB            & FLOAT     & Likelihood of BAL (not populated in this catalog) \\ 

68  & BI\_CIV              & FLOAT     & Equivalent width of  \ion{C}{IV} absorption by Balnicity index definition [$\mathrm{km\,s}^{-1}$]
 \\ 

69  & ERR\_BI\_CIV         & FLOAT     & Uncertainty in BI from \ion{C}{IV} absorption [km/s] \\ 

70  & NCIV\_2000           & INT32     & Number of absorption      troughs from \ion{C}{IV} extending $> 2000\,\mathrm{km\,s}^{-1}$ \\ 

71  & VMIN\_CIV\_2000      & FLOAT[5]  & Minimum velocity of BI absorption troughs blueward of \ion{C}{IV} [$\mathrm{km\,s}^{-1}$]\\ 

72  & VMAX\_CIV\_2000      & FLOAT[5]  & Maximum velocity of BI absorption troughs blueward of \ion{C}{IV} [$\mathrm{km\,s}^{-1}$]\\ 

73  & POSMIN\_CIV\_2000    & FLOAT[5]  & Velocity displacement from \ion{C}{IV} of normalized flux minimum for each AI absorption trough. [$\mathrm{km\,s}^{-1}$]\\
                                         
74  & FMIN\_CIV\_2000      & FLOAT[5]  & Value of normalized flux minimum for each BI absorption trough from \ion{C}{IV} \\ 

75  & AI\_CIV              & FLOAT     & Equivalent width of  \ion{C}{IV} absorption by absorption index definition [$\mathrm{km\,s}^{-1}$] \\ 

76  & ERR\_AI\_CIV         & FLOAT     & Uncertainty in AI from \ion{C}{IV} absorption [km/s]\\ 

77  & NCIV\_450            & INT32     & Number of absorption      troughs from \ion{C}{IV} extending $> 450\,\mathrm{km\,s}^{-1}$ \\ 

78  & VMIN\_CIV\_450       & FLOAT[17] & Minimum velocity of AI absorption troughs blueward of \ion{C}{IV} [$\mathrm{km\,s}^{-1}$]\\ 

79  & VMAX\_CIV\_450       & FLOAT[17] & Maximum velocity of AI absorption troughs blueward of \ion{C}{IV} [$\mathrm{km\,s}^{-1}$]\\ 

80  & POSMIN\_CIV\_450     & FLOAT[17] & Velocity displacement from \ion{C}{IV} of normalized flux minimum for each AI absorption trough [$\mathrm{km\,s}^{-1}$] \\ 
                                         
81  & FMIN\_CIV\_450       & FLOAT[17] & Value of normalized flux minimum for each AI absorption trough from \ion{C}{IV} \\ 
                                         
82  & BI\_SIIV             & FLOAT     & Equivalent width of  \ion{SI}{IV} absorption by Balnicity index definition [$\mathrm{km\,s}^{-1}$] \\ 

83  & ERR\_BI\_SIIV        & FLOAT     & Uncertainty in BI from \ion{Si}{IV} absorption [$\mathrm{km\,s}^{-1}$] \\

84  & NSIIV\_2000          & INT32     & Number of absorption      troughs from \ion{Si}{IV} extending $> 2000\,\mathrm{km\,s}^{-1}$ \\ 

85  & VMIN\_SIIV\_2000     & FLOAT[5]  & Minimum velocity of BI absorption troughs blueward of \ion{Si}{IV} [$\mathrm{km\,s}^{-1}$]\\ 

86  & VMAX\_SIIV\_2000     & FLOAT[5]  & Maximum velocity of BI absorption troughs blueward of \ion{Si}{IV} [$\mathrm{km\,s}^{-1}$]\\ 

87  & POSMIN\_SIIV\_2000   & FLOAT[5]  & Velocity displacement from \ion{Si}{IV} of normalized flux minimum for each BI absorption trough. [$\mathrm{km\,s}^{-1}$]\\

88  & FMIN\_SIIV\_2000     & FLOAT[5]  & Value of normalized flux minimum for each BI absorption trough from \ion{Si}{IV} \\ 

89  & AI\_SIIV             & FLOAT     & Equivalent width of  \ion{SI}{IV} absorption by absorption index definition [$\mathrm{km\,s}^{-1}$] \\ 

90  & ERR\_AI\_SIIV        & FLOAT     & Uncertainty in AI from \ion{Si}{IV} absorption [km/s]\\ 

91  & NSIIV\_450           & INT32     & Number of absorption  troughs from \ion{Si}{IV} extending $> 450\,\mathrm{km\,s}^{-1}$ \\ 

92  & VMIN\_SIIV\_450      & FLOAT[17] & Minimum velocity of AI absorption troughs blueward of \ion{Si}{IV} [$\mathrm{km\,s}^{-1}$]\\ 

93  & VMAX\_SIIV\_450      & FLOAT[17] & Maximum velocity of AI absorption troughs blueward of \ion{Si}{IV} [$\mathrm{km\,s}^{-1}$]\\ 

94  & POSMIN\_SIIV\_450    & FLOAT[17] & Velocity displacement from \ion{Si}{IV} of normalized flux minimum for each AI absorption trough. [$\mathrm{km\,s}^{-1}$]\\ 

95  & FMIN\_SIIV\_450      & FLOAT[17] & Value of normalized flux minimum for each AI absorption trough from \ion{Si}{IV} \\ 

96  & BALMASK              & BYTE      & Bitmask specifying whether BAL information was evaluated (see \S\ref{sec:data_model}) \\ 

103 & SNR\_CIV             & DOUBLE    & SNR in region extending from the \ion{C}{IV} emission feature to $25000\,\mathrm{km\,s}^{-1}$ bluewards\\ 

         \hline
    \end{tabular}
    \caption{Information about all of the BAL measurements included in the BAL catalog. More information about the columns is in the DESI EDR documentation for this Value Added Catalog and Section~\ref{sec:data_model}.}
    \label{tab:bal_cols}
\end{table*}

\subsection{Mock Spectra} \label{sec:mocks}

We test our BAL identification method using a set of simulated or mock quasar spectra. These mock spectra are generated with \texttt{quickquasars}, which is part of the DESI simulation package \texttt{desisim}\footnote{https://github.com/desihub/desisim} (see also \cite{herrera23}). The \texttt{quickquasars} code was designed to produce realistic DESI spectra in the DESI footprint with the same magnitude and redshift distribution as DESI quasars, as well as add realistic instrument noise and astrophysical systematics, such as BAL features. For quasars at sufficiently high redshift, the mock spectra also include Ly$\alpha$ forest absorption, damped Ly$\alpha$ absorbers, other high column density systems, and metal lines. The Ly$\alpha$ forest absorption is added with model transmission data generated with the \texttt{LyaColoRe} code described by \citet{farr20a}.\footnote{https://github.com/igmhub/LyaCoLoRe}

BALs are added into \texttt{quickquasars} with a series of empirical BAL templates developed by \citet{niu20}. The templates were developed from a subset of about $1500$ BAL quasars from SDSS DR14 catalogs during tests of a convolutional neural network BAL classifier by \citet{guo19}. Each template consists of the normalized absorption features as a function of velocity for a \ion{C}{IV} BAL detected in high SNR data after removal of the quasar continuum variations. The complete set of templates were selected to have the same AI and BI distributions as the complete DR14 BAL catalog. Each mock quasar has some probability of being a BAL (the default is 16\%) and if a quasar is a BAL, one of the templates is randomly assigned to that quasar. The BAL absorption features are added before the addition of noise or any Ly$\alpha$ and other absorption. The BAL absorption is applied to other lines with the same absorption profile as a function of velocity as the \ion{C}{IV} emission line. For each quasar that is assigned BAL absorption, \texttt{quickquasars} outputs a table of the BAL properties including the BAL template ID, the BAL redshift, and AI and BI values. We describe how we used these data to test the performance of the \texttt{balfinder} in Section~\ref{sec:cp}.

\section{Completeness and Purity} \label{sec:cp} 

\begin{figure}
    \includegraphics[width=\linewidth]{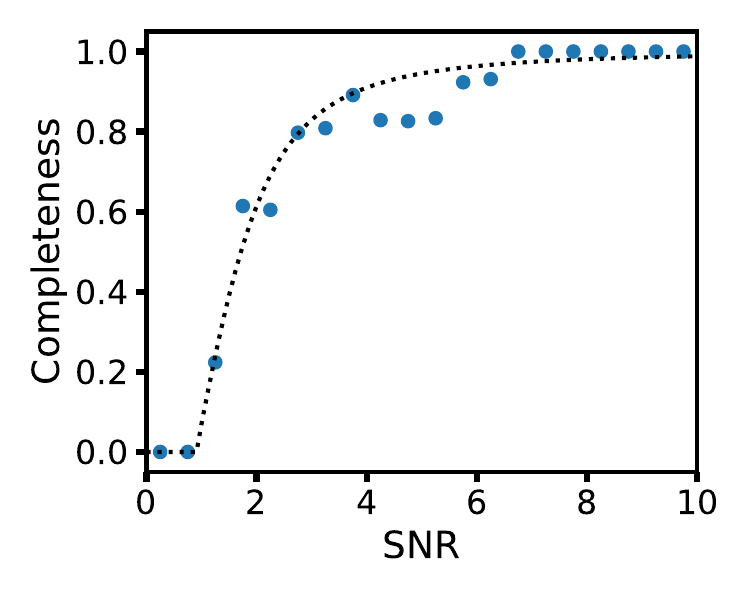}
    \captionsetup{width=\linewidth}
    \caption{Completeness of the \texttt{balfinder} from LyaCoLoRe mocks as a function of SNR. The completeness is shown in bins of 0.5 up to SNR$=10$. Relatively few spectra have higher SNR, and above this threshold the completeness asymptotes to 100\%. We use a simple, analytic approximation to the completeness vs.\, SNR ({\it dotted line}) to determine the average completeness of quasar catalogs with different SNR distributions. }
    \label{fig:performance}
\end{figure}


The completeness and purity of BAL catalogs are important to evaluate the impact of BALs on cosmological analysis. It is difficult to reliably quantify the completeness and purity of BAL identification algorithms on real BALs due to the challenge of constructing a useful truth catalog. Two previous studies by \citet{busca18} and \citet{guo19} relied on human-classified BALs for truth in order to test the performance of their classifiers and estimate what they referred to as the pseudo-completeness and pseudo-purity. The challenge of this approach is the reliability of human classification, which may have biases toward specific classes of BALs, redshifts, and/or SNR. In this paper, we instead quantify the performance with the DESI mocks described in Section~\ref{sec:mocks} that include simulated BAL spectra in 16\% of the quasars. We compare the \texttt{balfinder} output with the truth catalogs to determine the completeness and purity based on the $AI > 0$ criterion. 

Table \ref{tab:sum_props} contains estimates of the average completeness and purity of the $AI > 0$ BAL quasar sample. The average completeness and purity for SV1, meant to be representative of the full DESI data, is $56\%$ and $96\%$ respectively, with completeness being slightly lower for the lower average SNR catalogs (SV3 and M2). Although relatively low completeness is not detrimental to cosmological studies of large-scale structure with quasars, we recognize that applications such as quasar population studies, stacking, and variability require high completeness and purity. Those who wish to use this catalog for such applications may find the $BI > 0$ cut to be a more pure and complete sample, due to the more extreme absorption regions. This work did not estimate the $BI > 0$ completeness and purity using our mock spectra. Furthermore, masking objects with bad $\chi^2_{PCA}$ fits (i.e. >10) could also serve to further improve the purity of the sample. 

\subsection{Purity}



The purity is defined as the number of true BAL quasars recovered by the \texttt{balfinder} divided by the total number of BAL quasars identified by the \texttt{balfinder}. We use the AI criterion throughout, so that if the \texttt{balfinder} identifies any $AI > 0$ troughs, we consider this a BAL quasar. On average, we find that the purity of the \texttt{balfinder} is a little under $50\%$ for mock quasars. For comparison, \citet{busca13} measured a pseudo-purity of $77\%$ and \citet{guo19} measured a pseudo-purity of $40\%$. The significant difference relative to \citet{busca13} is because we use $AI > 0$ to define BALs, while they use a sample based on visual inspection that is biased towards BALs with stronger features. Our results are more similar to \citet{guo19} because they similarly used the $AI>0$ threshold to identify BALs. 

The $50\%$ pseudo-purity based on the mock spectra is not indicative of the purity we estimate for the BAL catalog presented in this work. This is because purity (and completeness) depend heavily on SNR. By binning the SNR of simulated BAL quasar spectra and evaluating the purity of each bin, we fit a simple model, similar to that shown in Figure \ref{fig:performance}. We find that the purity is very nearly $100\%$ for SNR < 11, but decreases rapidly to $50\%$ for SNR > 14. Using an estimation for the SNR blueward of the \ion{C}{IV} feature (\q{SNR\_CIV} column in the catalog, see Section \ref{sec:data_model}), we can then apply the purity function for each quasar to estimate the expected purity of the sample. From this, we estimate that the BAL quasar sample as a whole has a purity of $\sim99\%$. This seems reasonable since the average \q{SNR\_CIV} of the entire BAL sample is $\sim3$.

The decrease in purity for SNR > 11 is caused by the \texttt{balfinder} algorithm's failure to fit PCA components to high SNR spectra. This introduces a larger false-positive rate and drives down the purity. However, false positives are not a significant issue for cosmological analysis, as their main impact is that a small fraction of good data are thrown out, rather than the introduction of a source of contamination. This is also why we do not include a criterion based on the significance of the AI measurement, as the number of additional, masked pixels is very small. 


\subsection{Completeness}




The completeness is defined as the number of BAL quasars successfully recovered by the \texttt{balfinder} divided by the total number of true BAL quasars from the mocks. We again use the $AI > 0$ criterion and find the average completeness is about 68\% for the mock catalog. This is less than the 98\% pseudo-completeness reported by \citet{busca13} and the 97.4\% measured by \citet{guo19}. We attribute our lower completeness to both our $AI > 0$ criterion and the lower average SNR of our spectra. 

As with purity, we investigate the completeness as a function of SNR. The accompanying plot can be seen in Figure \ref{fig:performance}. We use the same method to estimate the completeness of the BAL quasar as purity, finding an average completeness of $\sim45\%$. Unlike the purity, the completeness is a very  strong function of SNR in the low-SNR regime ($\mathrm{SNR}<1$), while at higher SNR the completeness rapidly asymptotes to $\sim 95\%$ and is consequently comparable to the performance of previous algorithms. Unsurprisingly, it is harder for the algorithm to identify BAL quasars in noisy data, as noise spikes may interrupt a systematic depression of the quasar flux for the $450\,\mathrm{km\,s}^{-1}$ necessary to identify an $AI > 0$ BAL. 

Unlike previous work, our truth sample is not biased to only consider BALs in high SNR data, as it does not rely on visual inspection and the BAL features were added independent of the noise. As DESI quasar spectra will generally have lower SNR than SDSS spectra, especially for the tracer quasars at $z < 2.1$, we expect lower completeness. On the other hand, such lower completeness are not a significant cause for concern for cosmological studies, as the BAL features are less important relative to the noise properties of the data for those spectra. 


\section{DESI BAL Catalog} \label{sec:balcat} 

We construct the DESI EDR+M2 BAL catalog from quasar catalogs for SV1, SV3, and M2 as described in Section~\ref{sec:data}. While we conduct most of our analysis with the combined catalog, we also analyze them individually as they have somewhat different noise properties. The key difference is that the SV1 data are deeper and therefore higher SNR than what DESI plans to achieve at the end of the main survey, SV3 is representative of the four observations of $z>2.1$ quasars that DESI aims to achieve at the end of the main survey, and DESI-M2 mostly contains a single observation of each $z>2.1$ quasar and therefore is lowest in SNR. In the first subsection below we present the key properties of the BAL data model that is provided in the BAL catalog. The following subsection presents a summary of the BAL properties, and lastly we derive the BAL fractions for the SV1, SV3, and M2 datasets. 


\subsection{Data Model} \label{sec:data_model}

All of the DESI BAL catalogs have the same information as in a DESI quasar catalog plus additional columns with the BAL properties. There are a total of 39 additional columns. The 33 determined by the \texttt{balfinder} are summarized in Table~\ref{tab:bal_cols}. We add the remaining six after we mask the BAL features and refit the redshifts. Those additional columns are described below in Section~\ref{sec:balz}. Each of these properties is measured by the code provided in the public baltools repository. The remainder of this subsection explains each of the columns of the BAL catalog in more detail than is presented in Table~\ref{tab:bal_cols}. The online catalog also includes documentation.\footnote{https://data.desi.lbl.gov/doc/releases/edr/vac/balqso/} 

Columns 65 and 66 of the BAL catalog have information pertaining to the PCA fit used by the \texttt{balfinder} and discussed in Section~\ref{sec:identify}. The \q{PCA\_COEFFS} column specifies the coefficients for each of the five eigenspectra used to construct the PCA fit to each quasar \citep{guo19}, and the linear sum of these eigenspectra multiplied by the coefficients is the PCA fit to the spectrum. The column \q{PCA\_CHI2} is the reduced $\chi^2$ value of this fit. Any object in the catalog in the redshift range $1.57 < z < 5.0$ will have values in both of these columns, regardless of whether or not there are BAL features in the spectrum. 

Column 67 \q{BAL\_PROB} is not calculated in this implementation of the \texttt{balfinder}. This column is an artifact of the CNN approach used by \citet{guo19}. The value for each object for \q{BAL\_PROB} will be either zero or -99. A value of zero indicates that the object has been run through the \texttt{balfinder} code whereas a value of -99 means that BAL information was not evaluated for the object. The latter will indicate the quasar is outside of the redshift range in which the \texttt{balfinder} searches for BALs. We maintain this column for backwards compatibility and in the event we apply a CNN code or other classifier to future DESI quasar catalogs. 

Columns 68 through 95 record the strength and location of absorption BAL troughs in the quasar spectra. These values are calculated for the AI and BI criteria relative to both the \ion{C}{IV} and \ion{Si}{IV} lines. This includes a value for the absorption strength, errors for that value, the number of troughs, velocity ranges for each trough, and the location and depth of the minimum for each absorption feature. For each, a qualifier is specified for clarity in what the value corresponds to. For example, \q{\_2000} corresponds to values where the BI criterion was used and \q{\_450} corresponds to values where the AI criterion was used. Similarly, \q{\_SIIV} corresponds to features blueward of the \ion{Si}{IV} $\lambda$1398 line and the \q{\_CIV} indicator corresponds to features blueward of the \ion{C}{IV} $\lambda$1549 line. Since the \ion{Si}{IV} line is blueward of \ion{C}{IV}, BAL data for the \ion{Si}{IV} feature are not available for the lowest-redshift BALs. 

The column \q{BALMASK} is a bitmask that encodes information about any issues with running the \texttt{balfinder} on a given quasar. The definitions for \q{BALMASK} are that 0 means the \texttt{balfinder} ran successfully, 1 that the \texttt{balfinder} was not run, 2 that an object is outside of the BAL $z$ range, and 4 that the redshift difference between the input quasar catalog and the redshift after BAL masking are greater than 0.001. The bitmasks sum to specify any possible combinations of these cases. 

The \q{SNR\_CIV} column has the average SNR in the wavelength region where we search for BAL features associated with \ion{C}{IV}. We calculate the SNR of each pixel as the flux in a given pixel divided by the flux error provided by the spectroscopic pipeline \citep{guy23}. The average SNR is the average of all pixels that range from $v = 25000\,\mathrm{km\,s}^{-1}$ blueward of the \ion{C}{IV} $\lambda1549$ emission line to the line center at $v = 0$ that have $f(v) > 0$.

\subsection{Summary Properties} 

\begin{figure*}
    \includegraphics[width=\linewidth]{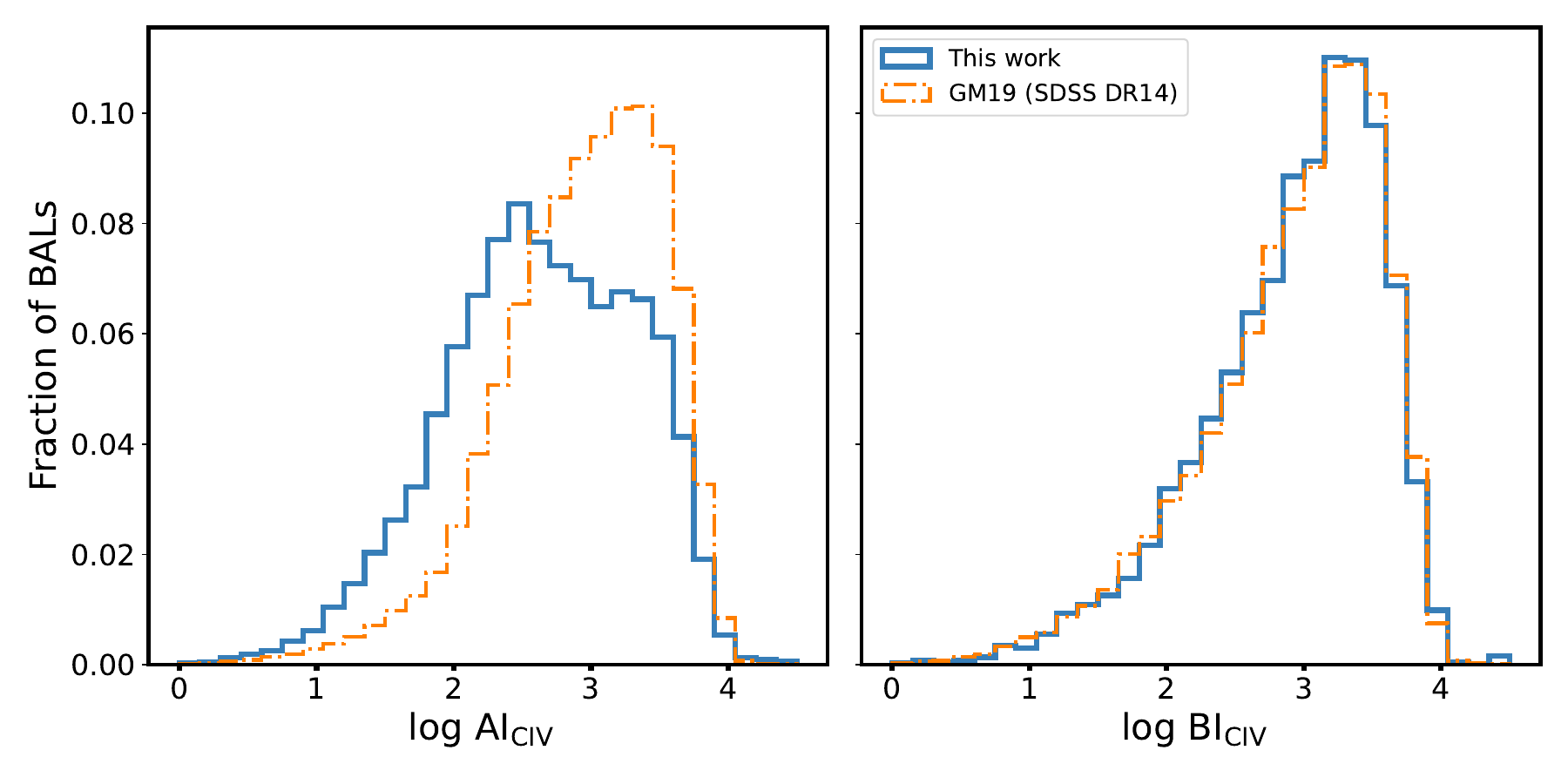}
    \caption{Logarithm of AI ({\it left panel}) and BI ({\it right panel}) values for BAL quasars. The distributions are normalized by the total number of quasars identified in each of the samples. The distributions for SDSS DR14 quasars from \citet{guo19} are shown for comparison. The normalizations are different for the AI and BI histograms (see Table~\ref{tab:sum_props}).} \label{fig:aibi_dist}
\end{figure*}

The BALs identified in this work are broadly similar to the BALs identified in earlier quasar catalogs from SDSS. For example, the AI and BI distributions of quasars in the DESI sample are qualitatively similar to the distributions in \cite{paris17} and \cite{trump06} which examined SDSS DR12 and SDSS DR3 quasars, respectively (see Figure~\ref{fig:aibi_dist}, {\it left}).  The AI distribution shows an excess of BALs at log(AI$_{\ion{C}{IV}}) <  3.3$, which is not as noticeable in the SDSS DR12 or DR14 quasars studied by \citet{paris17} and \citet{guo19}. This difference in the AI distributions is likely because the DESI catalogs use a simple $AI > 0$ criterion without consideration of the uncertainty in $AI$, while the earlier \citet{guo19} and SDSS catalogs used classifiers that rejected measurements of marginal significance. The BI distribution shown in the right panel of Figure~\ref{fig:aibi_dist} is in better agreement with previous work. This is  because BAL features need to be much more prominent to pass the BI criterion, and therefore this quantity is less impacted by a range of SNR values.  

We summarize a number of other important properties of the BAL catalogs in Table~\ref{tab:sum_props}. These include the total number of quasars in each catalog, the number in the redshift range where we can identify BALs, the number of quasars with one or more $AI > 0$ and $BI > 0$ troughs, and the percent of quasars in the redshift range that were identified as BAL quasars by each criterion. 

\begin{figure}
    \includegraphics[width=\linewidth]{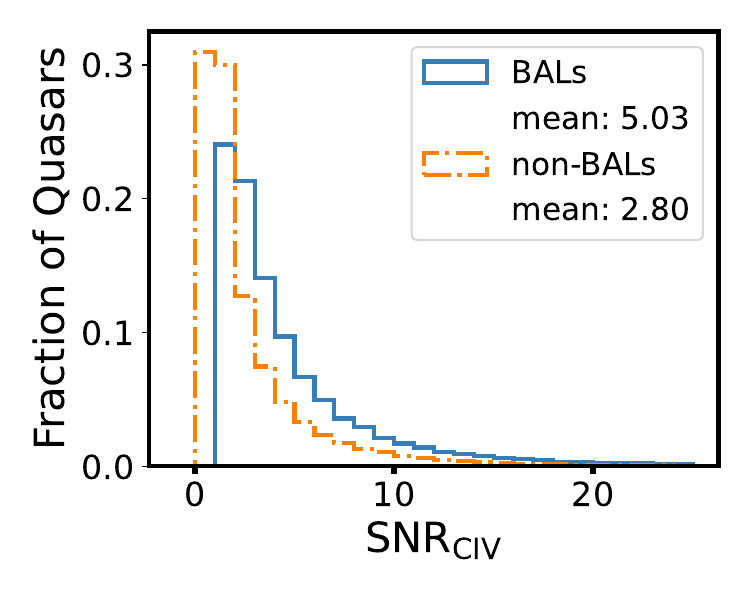}
    \captionsetup{width=3\linewidth}
    \caption{Distribution of SNR in the \ion{C}{IV} region for the BAL quasars  ({\it solid blue line}) and non-BAL quasars ({\it orange dotted line}). The SNR is calculated in the region where we search for absorption, which extends from the systemic wavelength of the \ion{C}{IV} emission to $25000\,\mathrm{km\,s}^{-1}$ blueward of the \ion{C}{IV} line. The SNR is the average of the product of the flux and the square root of the inverse variance for each pixel in this range. See \S\ref{sec:data_model} for more details.}
    \label{fig:snr_civ_dist}
\end{figure}

We also provide the average SNR of the quasars in the redshift range where we can identify BALs and an estimate of the completeness and purity of the BAL catalogs. The average SNR corresponds to all quasars in the range $1.57 < z < 5$, and not just those identified as BALs. We then use the SNR distribution for each catalog, combined with a simple function that matches the completeness and purity as a function of SNR presented in Section~\ref{sec:cp}, to calculate the average completeness and purity of each catalog. Figure~\ref{fig:snr_civ_dist} shows the SNR distribution of the BAL and non-BAL quasars. This is the SNR in the region blueward of \ion{C}{IV}. Note the mean SNR is higher for the BAL sample, which is expected based on our study of completeness. 


\begin{table*}
    \centering
    \begin{tabular}{p{1.25cm} p{1.5cm} p{2cm} p{1cm} p{1cm} p{1cm} p{1cm} p{1.75cm} p{1.75cm} p{1.25cm}}
         \multicolumn{9}{c}{DESI BAL Catalog Properties} \\
         \hline
         \hline 
         Catalog & Total QSOs & QSOs in z range & AI > 0 & AI \% & BI > 0 & BI \% & avg(SNR\_CIV) & Completeness & Purity\\

         \hline

         DESI SV1 & 37903 & 21486 & 4282 & 19.93 & 1001 & 4.66 &  4.41 & 0.56 & 0.96\\
         
         DESI SV3 & 50993 & 28641 & 5439 & 18.99 & 1338 & 4.67 & 3.99 & 0.51 & 0.97\\
        
         DESI-M2 & 280485 & 159772 & 20264 & 12.68 & 5058 & 3.17 & 2.44 & 0.42 & 0.99 \\
        
         All & 369381 & 209899 & 29985 & 14.29 & 7397 & 3.52 & 2.85 & 0.45 & 0.99\\

         \hline

         \hline
    \end{tabular}
    \caption{Statistical properties and numerical characteristics of each of the BAL catalogs. AI > 0 and BI > 0 refer to the number of objects identified with at least one absorption trough for each definition.}
    \label{tab:sum_props}
\end{table*}


\subsection{BAL Fraction Differences} 

As shown in Table~\ref{tab:sum_props}, the predicted completeness is highest for SV1, as it has the highest average SNR. This is because SV1 used longer exposure times than planned for DESI in order to obtain greater redshift completeness and thereby better quantify the performance of the target selection algorithms. SV1 also has the highest BAL fractions. The SV3 catalog is somewhat lower SNR than SV1, as the aim of SV3 was to obtain the same depth expected upon the completion of DESI. Lastly, the DESI-M2 main survey catalog has the lowest SNR as those spectra typically only have one observation of each Ly$\alpha$ quasar to date, while the goal is to obtain four or five observations of these quasars by the end of the survey. The completeness is consequently lowest for the main survey, and the BAL fractions are correspondingly lowest. We expect the main survey spectra will achieve the SV3 values by the end of the survey.

\section{Impact on Redshifts} \label{sec:balz}

The presence of BAL absorption features may impact the shape of the \ion{C}{IV}  and other emission lines, sometimes spectacularly (e.g., see Figure~\ref{fig:bal_spectra}), and produce redshift errors. In this section, we demonstrate that if we mask the locations of BAL features, we obtain better redshift measurements for the BAL quasars. The first subsection describes the masking procedure, and the second subsection describes the new redshift measurements and compares them to the values prior to masking the BAL features. The improvement in redshift has also been demonstrated with mock BAL spectra by \citet{angelagarcia23} and with observational data by \citet{brodzeller23}.



\subsection{Masking Procedure}\label{sec:masking}


Redshift estimation algorithms typically rely on some form of template fitting, and are especially susceptible to redshift errors if the observed quasar is not a great match to the templates. Due to the rich diversity of BALs in quasar spectra, it is not practical to account for all of this diversity with templates, and redshift errors are consequently more common in this subset of the quasar population. To reduce the bias and inherent redshift errors for the BAL sample, we adopted a procedure for masking out the wavelengths associated with the BAL features. 

Our approach starts with the BAL features cataloged on the blue side of the \ion{C}{IV} line and then assumes that absorption will also be associated with \ion{Si}{IV}, \ion{N}{V}, and Ly$\alpha$ at the same relative velocities. Absorption is sometimes present in these shorter-wavelength features, such as shown in the stacked BAL spectra studied by \citet{masribas19}, but this is not always the case. Consequently, the masking procedure implemented in this work is a conservative one, tending to over-mask the spectra before recomputing the redshift. These emission features are important for redshift measurements, as they are among the strongest emission lines in the observed wavelength range at high redshifts ($z>1.5$). 


The procedure used to mask the absorption features is relatively simple. Each object identified as a BAL quasar by the balfinder is assigned a \q{VMIN\_CIV\_450} and \q{VMAX\_CIV\_450} for each absorption feature. These numbers represent the velocity corresponding to an absorption feature up to $25000\,\mathrm{km\,s}^{-1}$ blueward of the \ion{C}{IV} emission line. For most quasars in the sample, only one absorption trough is present, although in extreme cases there may be many features. We mask these wavelengths by setting the inverse variance to zero for all of the wavelength values between each \q{VMIN\_CIV\_450} and \q{VMAX\_CIV\_450} pair from the BAL catalog, and repeat this for each of the emission lines considered. We make these changes in a copy of the original quasar spectral data, and the resulting file is identical to the original data except for the changes to the inverse variance values. 



\subsection{Masked BAL Redshifts}
\label{section:3.2}

Our next step is to rerun Redrock and measure new redshifts for the BAL quasars. The data used to create the quasar catalogs boasts a high completeness and purity for quasars, especially in the redshift regime for which BALs are characterized \citep{chaussidon23}. Because of this, we only fit the masked BAL quasar spectra with Redrock's quasar spectral templates, rather than all templates, as we are confident that the objects are quasars. We also assume that masked quasar redshifts will not significantly differ from the non-masked redshifts already available in the catalog. We therefore use the best (pre-masking) redshift to set a Gaussian prior for Redrock, with a width of 0.1.

The outputs from Redrock\footnote{See https://redrock.readthedocs.io/en/latest/api.html for more information} include the redshift \q{Z}, redshift error \q{ZERR}, a warning flag indicating the reliability of the redshift from Redrock \q{ZWARN}, the $\chi^2$ of the best-fit template \q{CHI2}, the difference between the best and second-best fit template \q{DELTACHI2}, and the spectral classification \q{SPECTYPE} (S. Bailey et al. (2024), in preperation). These data are appended to the BAL catalog with the extra qualifier \q{\_MASK} to make clear that these data come from the run of Redrock after BAL features are masked. Objects which are not BAL quasars will have nonsensical values for all \q{\_MASK} column values, for example, \q{Z\_MASK} for a non BAL quasar is set to -99. \q{SPECTYPE\_MASK} is always \q{QSO} for BAL quasars since only quasar templates are used to rerun Redrock after BAL features are masked.

\begin{figure*}
    \includegraphics[width=\linewidth]{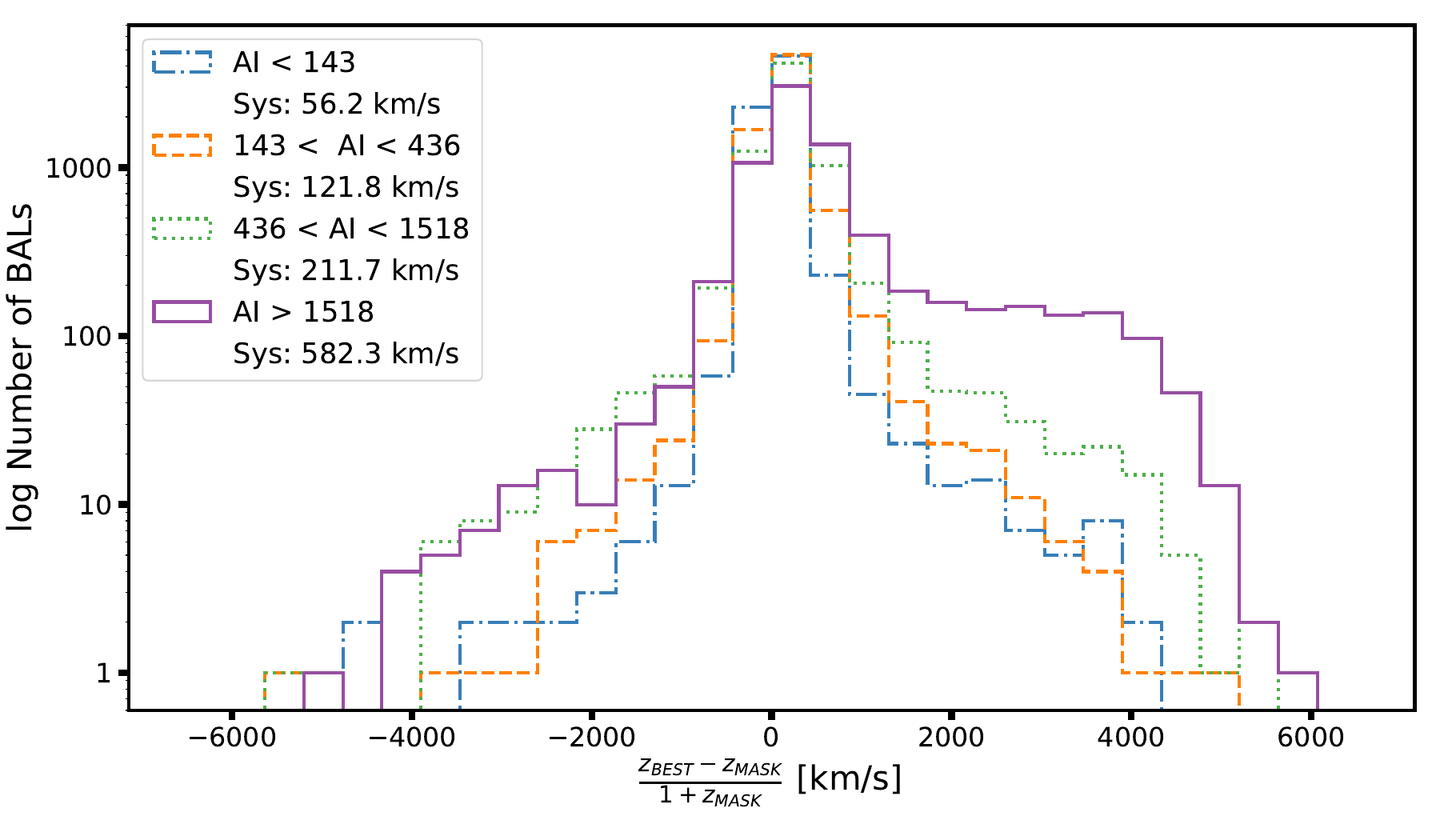}
    \captionsetup{width=\linewidth}
    \caption{Distribution of velocity differences between the original pipeline redshift and the new redshift after masking BAL features. The distributions are binned into quartiles by AI value. The mean difference is a strong function of AI value in the sense that the velocity different is relatively modest for weak BALs (smallest AI quartile) and increases to $582.3\,\mathrm{km s}^{-1}$ for the greatest quartile, which shows redshift errors are more significant for larger AI values. In addition, the velocity differences are strongly skewed toward positive velocity differences, which correspond to a systematic underestimate of the true redshift before masking. This is due to the absorption on the blue side of the line.}
    \label{fig:dv_dist}
\end{figure*}

The velocity differences c$(z_{nomask} - z_{mask})/(1+z_{mask})$ before and after masking range from on order $-1000$ to $+1000\,\mathrm{km\,s}^{-1}$, with the average skewed toward positive values. This offset means that the redshifts after masking are lower than the redshifts before masking, which is consistent with our expectation that the BAL absorption has removed flux from the blue side of the emission lines and biases the line centers redward of their true location. Figure~\ref{fig:dv_dist} shows histograms of the velocity differences. The velocity difference averaged over all of the BALs is $243\,\mathrm{km\,s}^{-1}$. 

We separated the BAL sample into four quartiles of AI to determine if there is a trend in the size of the velocity shift with the strength of the absorption. This trend is significant and is shown in Figure~\ref{fig:dv_dist}. The shift ranges from an average velocity shift of $56.2\,\mathrm{km\,s}^{-1}$ for the first quartile of AI to $582.3\,\mathrm{km\,s}^{-1}$ for the fourth quartile. The tail of the distribution also becomes longer when more highly absorbed objects are in the sample. This result is expected, as more absorption in the quasar spectra corresponds to a larger portion of the spectra with a low flux contribution, and this causes the fits to misinterpret the peak of emission line features. 


In addition to this systematic shift in the velocity, we found that 6.7\% of the BAL quasars in the sample have changes in redshift greater than the catastrophic error regime required for tracer quasars in DESI ($\Delta v > 1000\mathrm{km\,s}^{-1}$) whereas 9.3\% have redshift changes greater than the cutoff required for random errors \citep[$\Delta v > 750\,\mathrm{km\,s}^{-1}$][]{svpaper} .



\section{Prospects for Rarer BAL Classes in DESI} \label{sec:unusualbals}

\begin{figure*}
    \includegraphics[width=\linewidth]{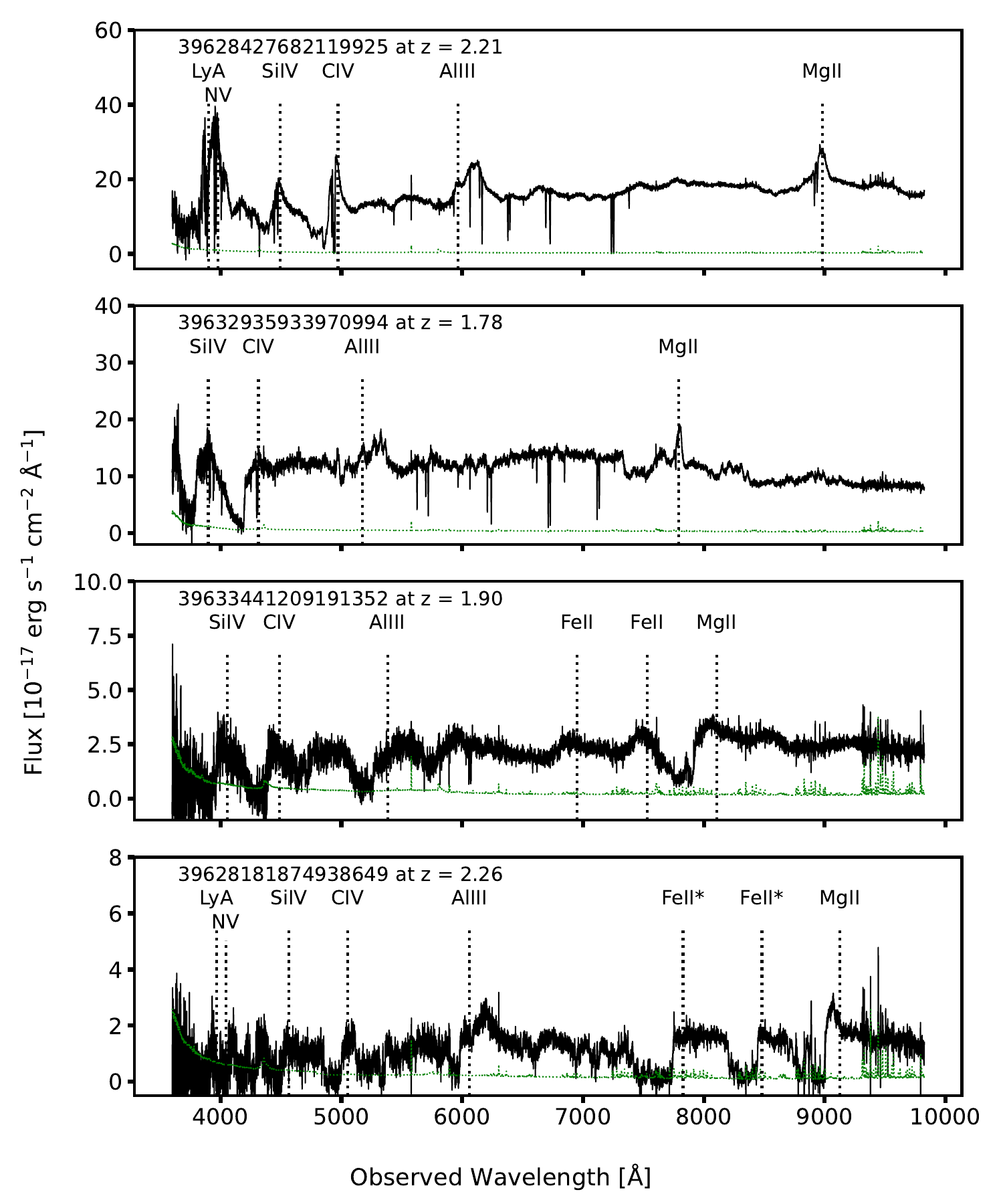}
    \captionsetup{width=\linewidth}
    \caption{Examples of BALs from the DESI EDR. The top panel shows a BAL at $z=2.21$ dominated by high-ionization troughs associated with \ion{Al}{III}, \ion{C}{IV}, and {Si}{IV}, although there is weak evidence for some broad, highly blueshifted absorption associated with \ion{Mg}{II}. The second panel from the top shows a BAL at $z=1.78$ with the broad \ion{Mg}{II} absorption characteristic of LoBAL quasars, in addition to the common high-ionization features. The third panel shows a BAL at $z=1.90$ with weak Fe absorption, most notably \ion{Fe}{II} absorption, that is characteristic of FeLoBALs. The bottom panel shows a BAL at $z=2.26$ with the very strong Fe absorption characteristic of FeLoBALs. The flux uncertainty is also shown in each panel ({\it green, dotted line}). }
    \label{fig:unusualbals}
\end{figure*}

The large samples of BAL quasars from spectroscopic surveys like SDSS and DESI offer a great opportunity to identify larger samples of rare BALs \citep[e.g.,][]{hall02}, and the frequency and properties of such systems may lead to new insights into quasar physics. In a study of quasars in the SDSS Third Data Release (DR3), \citet{trump06} measured the incidence of HiBALs, LoBALs and FeLoBALs. They found 26\% of quasars are HiBALs with the $AI>0$ criterion in the redshift range $1.7 \leq z \leq 4.38$ where they could detect \ion{C}{IV} absorption. The rate for LoBALs was much lower at 1.31\% in the redshift range $0.5 \leq z \leq 2.15$, and the rate for FeLoBALs was lower still at about 0.33\%. Many of the rarer classes of BALs were identified through visual inspection, especially the FeLoBALs, as the absorption may be so significant that automated techniques fail. A further complication is that BAL quasars, especially LoBALs and FeLoBALs, tend to have greater reddening than non-BAL quasars \citep[e.g.][]{hall97,najita00,morabito19}. 

\citet{alexander23} performed visual inspection of a large number of quasars from DESI Survey Validation and identified many BAL quasars. We have performed a complimentary visual inspection of BALs identified with the \texttt{balfinder} that exhibited poor fits. The example spectra shown in Figure~\ref{fig:unusualbals} include ones dominated by high-ionization features of \ion{Al}{III}, \ion{C}{IV}, and {Si}{IV} ({\it top panel}), and below that examples of a LoBAL quasar with \ion{Mg}{II} absorption and two FeLoBALs with prominent absorption that is blueshifted relative to the \ion{Fe}{II} multiplets near 2400\,\AA\ and 2600\,\AA. DESI will observe approximately two million quasars in the redshift range where it could detect broad absorption lines associated with \ion{Mg}{II} and \ion{Fe}{II}, and therefore has the potential to identify several tens of thousands of LoBALs and several thousand FeLoBALs. This could include extremely rare objects like the one studied by \citet{choi20}. 

Many studies of BALs have used their variability to study their physical properties. For example, \citet{filiz14} measured changes in the \ion{C}{IV}, \ion{Si}{IV}, \ion{Al}{III} features and compared correlations to disk wind models. One such correlation is that lines of sight with more low ionization material also exhibit broader and deeper \ion{C}{IV} troughs. \citet{grier16} used multi-epoch observations from several SDSS programs and found that the vast majority show no evidence of acceleration, which implies most of the absorbing material is at large distances from the black hole and/or is not decelerating due to collisions with ambient material along the line of sight. \citet{mcgraw15} studied the absorption line variability of the subclass of FeLoBAL quasars and concluded that the absorbing material can be up to tens of parsecs from the central source. DESI will obtain multiple observations of quasars at $z > 2$ to improve the signal-to-noise ratio of the Ly$\alpha$ sample. This will enable variability studies of a significant subset of the many thousands of LoBAL and FeLoBAL quasars that DESI will discover. 

Even though the vast majority of DESI spectra will be lower SNR than targeted follow-up studies of BAL quasars, these data will be valuable for stacking analyses. \citet{hamann19} created composite spectra based on \ion{C}{IV} strength and \ion{Al}{III} to investigate weaker features, including ones normally compromised by Ly$\alpha$ forest absorption. They found interesting constraints on the typical densities and distances of the absorbing material, and that the LoBALs tend to have the largest outflow column densities and highest velocities. In another application of stacked BAL spectra, \citet{masribas19} found evidence of radiative acceleration through line locking in the \ion{C}{IV} line. High SNR stacks of DESI BALs or of individual, rare objects could lead to further insights with the continued growth in sophistication of analysis tools, such as the SimBAL analysis code \citep{leighly22,choi22a,choi22b}. 

\section{Summary} \label{sec:summary}

The DESI BAL catalogs together comprise one of the largest samples of BAL quasars to date, with just under 30,000 BAL quasars identified by our automated searching algorithm. In this paper, we describe the algorithm employed by the DESI collaboration to identify and measure the properties of BAL quasars. This algorithm identifies BALs via absorption on the blue side of the \ion{C}{IV} emission feature, and therefore we only identify BALs in quasars between $1.57 < z < 5$ when this spectral region is clearly visible by DESI. This restriction means that all of the BAL quasars appear to be HiBALs, although we have not conducted a similar, systematic search in the vicinity of low-ionization features such as \ion{Mg}{II}. We record the number of absorption troughs that satisfy both the AI and BI criteria and find the distribution of both quantities is similar to previous results from SDSS, except that DESI is somewhat more sensitive to lower AI values. We also record the velocity limits of each absorption trough. The limits of each absorption feature are important to mask spectral regions that may be contaminated by BAL absorption. 

We used mock catalogs to analyze the completeness and purity of our algorithm. These mock catalogs use a range of empirical templates derived from high SNR BAL spectra from SDSS that are randomly assigned to the mock quasars. We find that the completeness of our algorithm depends on SNR and decreases significantly for $\mathrm{SNR} < 2$ per pixel. This dependence implies that the completeness will vary between surveys, and we see a difference in the BAL fraction for different DESI catalogs that is consistent with the average completeness of the spectra in each catalog. The One-Percent Survey was designed to be representative of the final DESI dataset. Our results from the SV3 catalog suggest that the expected fraction of BAL quasars identified by the \texttt{balfinder} with at least one AI absorption trough should be $\sim19$\% once DESI reaches full survey depth. 

Historically, one major motivation to identify BALs in cosmological surveys was to eliminate them as a potential source of contamination. One rationale was that BAL features may produce greater redshift errors, which negatively impacts quasar clustering. Another was that BAL absorption features in the Ly$\alpha$ forest region may contaminate measurements of the neutral Hydrogen distribution in quasars above $z>2$.  In order to mitigate the impact of BALs on quasar redshift errors, we mask the BAL features  associated with the \ion{C}{IV}, \ion{Si}{IV}, \ion{N}{V}, and Ly$\alpha$ lines and report new redshifts for these quasars. Mock quasar studies by \citet{angelagarcia23} show that masking the BAL features decreases the redshift errors to be comparable to those for non-BAL quasars. We show that the velocity shifts of quasars before and after masking are correlated with the strength of absorption and that masking the BAL troughs shifts the redshifts somewhat to the blue. In some cases, the velocity shifts after masking are as large as thousands of $\mathrm{km\,s}^{-1}$, which are greater than the $1000\,\mathrm{km\,s}^{-1}$ criterion that DESI  has adopted to define catastrophic redshift errors, although the average velocity shift is $\sim243\,\mathrm{km\,s}^{-1}$.

Keeping BAL quasars in DESI increases the statistical power for quasar clustering measurements and increases the number of sightlines that probe the Ly$\alpha$ forest. The projected BAL quasar fraction of $\sim19$\% identified by the \texttt{balfinder} will significantly impact the number of sightlines ultimately kept for the cosmological analysis. Specifically, \citet{ennesser22} showed with eBOSS observations that adding back in the 12\% of BAL quasars in eBOSS decreased the uncertainties on the autocorrelation function by 12\%. While some of the spectral range is masked in BALs, this is offset by the higher average SNR of each BAL quasar relative to non-BAL quasars, or at least relative to quasars where we do not detect BAL features. In the DESI sample, the average SNR per pixel of BALs is about five, as opposed to just under three for non-BAL quasars. This and future BAL catalogs will be an invaluable part of maximizing the scientific return of the DESI survey. 


\section*{Acknowledgements}

SMF is grateful for support from Ohio State that includes two College of Arts and Sciences Undergraduate Research Scholarships, summer research support from the Departments of Astronomy and Physics, and scholarships including the Smith Student Support scholarship and the Ann Slusher Tuttle undergraduate scholarship.
PM and LE acknowledges support from the United States Department of Energy, Office of High Energy Physics under Award Number DE-SC-0011726. 

This material is based upon work supported by the U.S. Department of Energy (DOE), Office of Science, Office of High-Energy Physics, under Contract No. DE–AC02–05CH11231, and by the National Energy Research Scientific Computing Center, a DOE Office of Science User Facility under the same contract. Additional support for DESI was provided by the U.S. National Science Foundation (NSF), Division of Astronomical Sciences under Contract No. AST-0950945 to the NSF’s National Optical-Infrared Astronomy Research Laboratory; the Science and Technology Facilities Council of the United Kingdom; the Gordon and Betty Moore Foundation; the Heising-Simons Foundation; the French Alternative Energies and Atomic Energy Commission (CEA); the National Council of Science and Technology of Mexico (CONACYT); the Ministry of Science and Innovation of Spain (MICINN), and by the DESI Member Institutions: \url{https://www.desi.lbl.gov/collaborating-institutions}. Any opinions, findings, and conclusions or recommendations expressed in this material are those of the author(s) and do not necessarily reflect the views of the U. S. National Science Foundation, the U. S. Department of Energy, or any of the listed funding agencies.

The authors are honored to be permitted to conduct scientific research on Iolkam Du’ag (Kitt Peak), a mountain with particular significance to the Tohono O’odham Nation.

\section*{Data Availability}

The BAL catalogs for data collected during DESI EDR observations can be downloaded from
https://data.desi.lbl.gov/doc/releases/edr/vac/balqso/.
Data points for the Figures in this publication can be accessed in the following zenodo
repository https://zenodo.org/record/8267633.


\bibliographystyle{mnras}
\bibliography{ref} 






\bsp	
\label{lastpage}
\end{document}